\documentclass[aps,prb,twocolumn,showpacs,superscriptaddress,floatfix]{revtex4-1}

\usepackage{graphicx,graphics}
\usepackage{dcolumn}
\usepackage{amsmath,amssymb,amsfonts}
\usepackage{latexsym,verbatim}
\usepackage{bm}
\usepackage{color}
\usepackage{ulem}
\usepackage[breaklinks=true,colorlinks,citecolor=blue,linkcolor=blue,urlcolor=blue]{hyperref}
\usepackage[percent]{overpic}
\usepackage{bbold}
\usepackage{amsmath}
\usepackage{footnote}
\usepackage{comment}
\usepackage{epsfig}

\begin{document}

\title{Resonant tunneling and the quasiparticle lifetime in graphene/boron nitride/graphene heterostructures}

\author{Karina A. Guerrero-Becerra}
\affiliation{CNR-NANO Research Center S3, Via Campi 213/a, 41125 Modena, Italy}

\author{Andrea Tomadin}
\affiliation{NEST, Istituto Nanoscienze-CNR and Scuola Normale Superiore, I-56126 Pisa, Italy}

\author{Marco Polini}
\affiliation{Istituto Italiano di Tecnologia, Graphene Labs, Via Morego 30, I-16163 Genova,~Italy}

\begin{abstract}
Tunneling of quasiparticles between two nearly-aligned graphene sheets 
produces resonant current-voltage characteristics because of the quasi-exact conservation of in-plane momentum. We claim that, in this regime, vertical transport in graphene/boron nitride/graphene heterostructures carries precious information on electron-electron interactions and the quasiparticle spectral function of the two-dimensional electron system in graphene. We present extensive microscopic calculations of the tunneling spectra with the inclusion of quasiparticle lifetime effects and elucidate the range of parameters (inter-layer bias, temperature, twist angle, and gate voltage) under which electron-electron interaction physics emerges.
\end{abstract}

\maketitle

\section{Introduction}
\label{sect:intro}

The quantum lifetime~\cite{mahan_book_1981} of electrons roaming in semiconductors and semimetals is the result of microscopic scattering events between electrons and disorder, lattice vibrations, and other electrons in the Fermi sea. 

At low temperatures, the lifetime of electrons close to the Fermi surface is dominated by elastic scattering off of the
static disorder potential in the material. Upon increasing temperature, however, inelastic scattering mechanisms like electron-phonon and electron-electron (e-e) scattering begin to play a role. Standard electrical transport measurements are sensitive to elastic scattering and electron-phonon processes. The e-e scattering time $\tau_{\rm ee}$, which in a normal Fermi liquid coincides with the quasiparticle lifetime~\cite{Nozieres,Pines_and_Nozieres,Giuliani_and_Vignale}, is much harder to extract from dc transport since such e-e scattering processes conserve the total momentum of the electron system. At low temperatures, order-of-magnitude estimates of $\tau_{\rm ee}$ are often obtained from weak localization measurements~\cite{altshuler_jpc_1982,imry_sst_1994} of the dephasing time $\tau_\phi$.

Direct measurements of $\tau_{\rm ee}$ are however possible. Any experiment that accesses the so-called quasiparticle spectral function~\cite{Nozieres,Pines_and_Nozieres,Giuliani_and_Vignale} ${\cal A}({\bm k},\varepsilon; \mu)$, is sensitive to $\tau_{\rm ee}$.
(Here, ${\bm k}$, $\varepsilon$, and $\mu$ denote wave vector, energy, and chemical potential, respectively.)
It is well known that angle-resolved photoemission spectroscopy (ARPES)~\cite{damascelli_rmp_2003} is one of such experiments. In the case of graphene, accurate ARPES measurements~\cite{bostwick_naturephys_2007,zhou_naturemater_2007,bostwick_science_2010,walter_prb_2011,siegel_pnas_2011} of ${\cal A}({\bm k},\varepsilon; \mu)$ require large flakes and have therefore been limited to high-quality epitaxial samples grown on the silicon or carbon face of SiC. 

What is less known is that tunneling between two two-dimensional (2D) electron systems~\cite{murphy_prb_1995,Giuliani_and_Vignale} with {\it simultaneous} conservation of energy and momentum also probes ${\cal A}({\bm k},\varepsilon; \mu)$ and therefore $\tau_{\rm ee}$. In these experiments, the tunnel current flowing perpendicularly between two parallel 2D electron systems separated by a barrier is measured.  The conservation of in-plane momentum ${\bm k}$ strongly constrains the phase space for tunneling processes and grants unique access to the quasiparticle spectral function ${\cal A}({\bm k},\varepsilon; \mu)$. 2D-to-2D tunneling spectroscopy was carried out by Murphy {\it al.}~\cite{murphy_prb_1995} on double quantum well heterostructures consisting of two GaAs quantum wells separated by an undoped ${\rm Al}_x{\rm Ga}_{1-x}{\rm As}$ barrier with a width $d$ in the range $17.5~{\rm nm} \leq d \leq 34~{\rm nm}$. Experimental results for the width of the tunneling resonances were compared with available theoretical results on the quasiparticle lifetime of a 2D parabolic-band electron system~\cite{chaplik_jetp_1971,hodges_prb_1971,fukuyama_prb_1983,giuliani_prb_1982} and stimulated much more theoretical work~\cite{jungwirth_prb_1996,zheng_prb_1996,reizer_prb_1997,marinescu_prb_2002,qian_prb_2005}.

\begin{figure}[t]
\begin{overpic}[width=\linewidth]{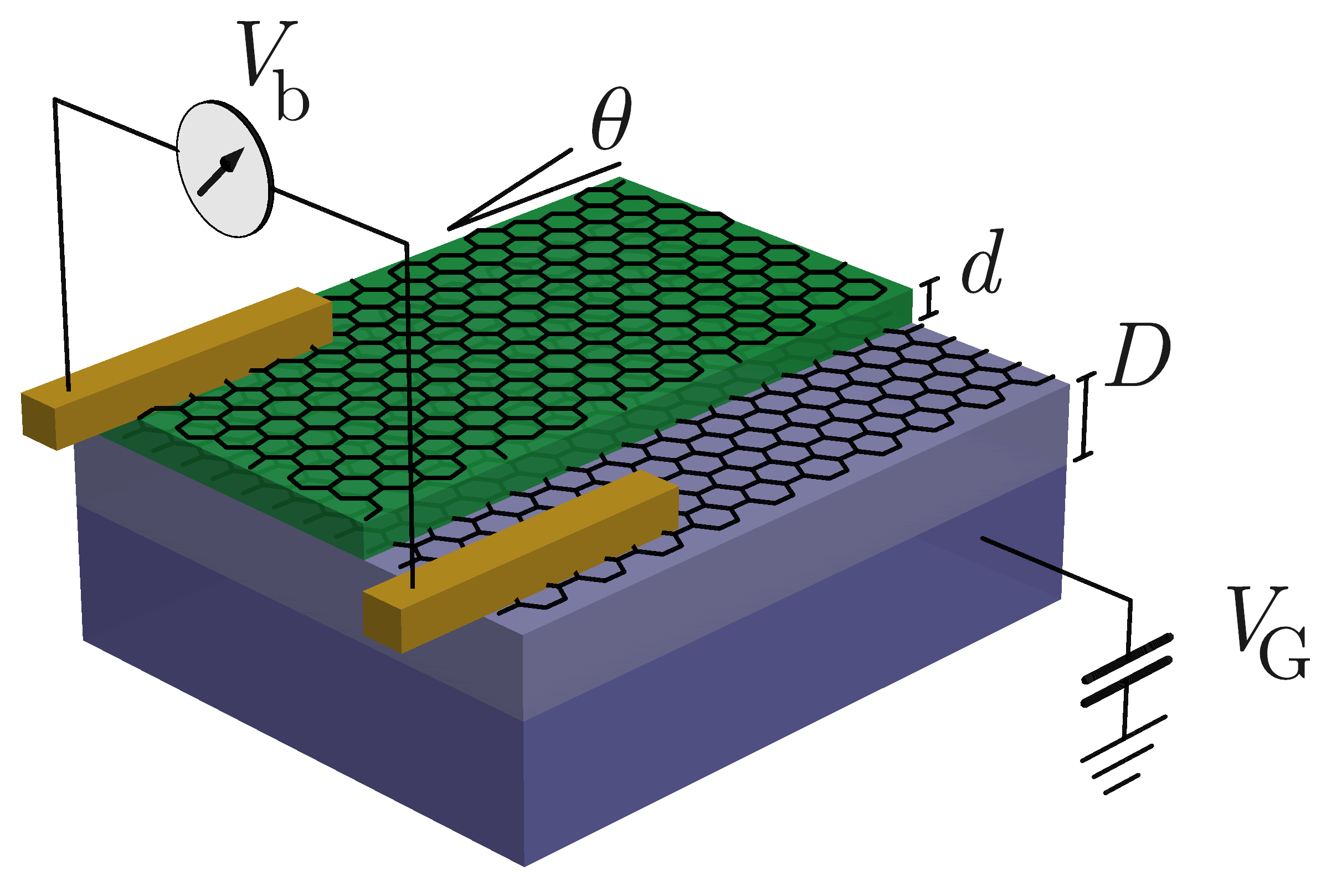}\end{overpic}
\caption{\label{fig:setup}
(Color online) Pictorial representation of the tunneling heterostructure considered in this Article. 
From bottom to top, it includes a back gate maintained at the electric potential $V_{\rm G}$ (purple), an insulating slab of thickness $D$, a bottom graphene layer, a hBN slab of thickness $d$ (green), and a top graphene layer.
The misalignment angle between the two graphene layers is $\theta$.
Ohmic contacts (gold) are deposited on the two graphene layers and an inter-layer bias $V_{\rm b}$ is applied between the top and bottom graphene layers.}
\end{figure}

Recently, a large number of 2D-to-2D tunneling spectroscopy experiments has been carried out in van der Waals heterostructures~\cite{geim_nature_2013} comprising two graphene sheets separated by a hexagonal boron nitride (hBN) barrier~\cite{britnell_nanolett_2012,britnell_science_2012,britnell_naturecommun_2013,mishchenko_naturenano_2014}. In particular, this work is motivated by the recently gained ability to {\it align} the two graphene crystals~\cite{britnell_naturecommun_2013,mishchenko_naturenano_2014,fallahazad_nanolett_2015}, 
which enables  tunneling measurements in which the in-plane momentum ${\bm k}$ is nearly exactly conserved. Here, we present a theoretical analysis of the role of {\it intra-layer} e-e interactions on the tunneling characteristics of graphene/hBN/graphene heterostructures as sketched in Fig.~\ref{fig:setup}.
To the best of our knowledge, all available theoretical studies~\cite{feenstra_jourapplphys_2012,delabarrera_jvactechnol_2014,brey_prapp_2014,wallbank_thesis_2014,mishchenko_naturenano_2014,lane_apl_2015} of tunneling in these heterostructures have not dealt with e-e interaction effects.

Our Article is organized as following. In Sect.~\ref{sect:hamiltonian} we present the tunneling Hamiltonian and an expression for the tunneling current $I=I(V_{\rm b})$ as a functional of the quasiparticle spectral function ${\cal A}_{\lambda}({\bm k}, \varepsilon; \mu)$ for conduction- ($\lambda = +$) and valence-band ($\lambda = -$) states in each layer. In Sect.~\ref{sect:details} we present two crucial ingredients for the calculation of the tunneling current: i) electrostatic relations linking the chemical potentials $\mu_{\rm T}$ and $\mu_{\rm B}$ in the two layers with gate voltage $V_{\rm G}$ and inter-layer bias $V_{\rm b}$ and ii) the quasiparticle spectral function $A_{\lambda}(\bm{k},\varepsilon; \mu)$ with the inclusion of quasiparticle lifetime effects. In Sect.~\ref{sect:numerics} we present and discuss our main numerical results. Finally, in Sect.~\ref{sect:summary} we summarize our main findings.

\section{Tunneling Hamiltonian and current-voltage characteristics}
\label{sect:hamiltonian}

We consider the setup depicted in Fig.~\ref{fig:setup}, consisting of two parallel graphene layers, separated by a tunneling barrier of thickness $d$. The misalignment angle between the lattices of the two graphene layers is denoted by $\theta$. The bottom layer is separated from a back gate by an insulating layer of thickness $D$.
The back gate is maintained at the electric potential $V_{\rm G}$ while an electric potential bias $V_{\rm b}$ is applied between the top and bottom graphene layers.
Our aim is to calculate the tunneling current density $I$ between the two layers, as a function of the applied bias $V_{\rm b}$. (The total tunneling current is obtained by multiplying the current density by the area of the region where the two graphene layers overlap.)

We model the tunneling heterostructure in Fig.~\ref{fig:setup} with the following Hamiltonian in the layer-pseudospin basis:
\begin{equation} \label{eq:eff-ham}
\hat{\cal H}^{\rm eff}  =   
\begin{pmatrix}
\hat{\cal H}_{\rm T} &  0 \\
0 & \hat{\cal H}_{\rm B} \\ 
\end{pmatrix}+
\begin{pmatrix}
0 &  \hat{\cal H}_{\rm TB} \\
\hat{\cal H}_{\rm TB}^{\dagger} & 0 \\ 
\end{pmatrix}~.
\end{equation}
Here, $\hat{\cal H}_{\rm T}$ ($\hat{\cal H}_{\rm B}$) is the 2D massless Dirac fermion 
Hamiltonian~\cite{kotov_rmp_2012} of the top (bottom) graphene layer and
\begin{equation} \label{eq:tunnham}
\hat{\cal H}_{\rm TB}  =  \frac{\gamma_{\rm eff}}{3} \sum_{j = 1,2,3} e^{-i \Delta {\bm K}_{j} \cdot \hat{\bm r}} 
\begin{pmatrix}
1 & e^{-i \frac{2\pi}{3}(j -1)} \\
e^{i \frac{2\pi}{3}(j -1)} & 1 \\ 
\end{pmatrix}
\end{equation}
is the tunneling Hamiltonian between two graphene layers in the lattice-pseudospin basis~\cite{bistrizer_prb_2010,bistrizer_pnas_2011,bistrizer_prb_2011,mele_prb_2011,dossantos_prb_2012}. Eq.~(\ref{eq:tunnham}) assumes that: 1) tunneling between the two graphene layers occurs through highly misaligned hBN~\cite{mishchenko_naturenano_2014} (which is therefore treated as a homogeneous dielectric); 2) chirality of the eigenstates of the 2D massless Dirac fermion Hamiltonians $\hat{\cal H}_{\rm T}$ and $\hat{\cal H}_{\rm B}$ is preserved upon tunneling~\cite{mishchenko_naturenano_2014}; and 3) {\it inter-layer} e-e interactions are negligible. While assumptions 1) and 2) are certainly reasonably justified, assumption 3) is certainly unjustified since tunnel experiments in graphene/hBN/graphene heterostructures~\cite{britnell_nanolett_2012,britnell_science_2012,britnell_naturecommun_2013,mishchenko_naturenano_2014} are always carried out in the strong coupling regime~\cite{gorbachev_naturephys_2012,carrega_njp_2012}, i.e.~$d k_{\rm F, T(B)} \ll 1$, where $k_{\rm F, T(B)}$ is the Fermi wave vector in the top (bottom) graphene layer. This is at odds with aforementioned tunneling experiments in (and related theory work on) double quantum well heterostructures consisting of two GaAs quantum wells separated by undoped ${\rm Al}_x{\rm Ga}_{1-x}{\rm As}$ barriers. Relaxing assumption 3) is certainly an interesting conceptual endavor, which is well beyond the scope of the present Article and is left for future work.

In Eq.~(\ref{eq:tunnham}),  $\gamma_{\rm eff}$ is an effective inter-layer coupling strength, which strongly depends on the thickness $d$ of the hBN barrier, and $\Delta \bm{K}_{j}\equiv \theta \hat{\bm z} \times \bm{K}_{j}$. 
Here, ${\bm K}_{j}$ with $j=1,2,3$ denote the three equivalent positions of the corners of the Brillouin zone of the bottom layer, with $\theta=0$ denoting the A-A stacking configuration.
Physically, the quantity $\Delta \bm{K}_{j}$ ($-\Delta \bm{K}_{j}$) represents the in-plane wave vector change of electrons upon tunneling from the top to the bottom (bottom to the top) layer. The matrix elements of the Hamiltonian~(\ref{eq:tunnham}) between plane-wave states in the two different layers read as following:
\begin{eqnarray} \label{eq:matrixelements}
t_{\lambda, \lambda'}(\bm{k},\bm{k}') & = & 
\gamma_{\rm eff} \frac{(2 \pi)^2}{6} \sum_{j= 1,2,3} \left\{1+ \lambda e^{-i [2 \pi(j-1)/3 - \varphi_{\bm k}]}\right\} \nonumber \\
&& \times \left\{1+\lambda' e^{i [2 \pi(j-1)/3 - \varphi_{{\bm k}'}]}\right\}~,
\end{eqnarray}
where $\lambda, \lambda' = \pm$ are band indices, ${\bm k}$ (${\bm k}'$) is the wave vector of the electronic  state in the bottom (top) layer, with polar angle $\varphi_{\bm k}$ ($\varphi_{{\bm k}'}$). 

To second order in the inter-layer coupling $\gamma_{\rm eff}$, the tunneling current density is given by~\cite{mahan_book_1981,wolf_book_2012}

\begin{eqnarray}\label{eq2:current}
I(V_{\rm b}) & = & \frac{e}{2\pi h} N_{\rm f} \sum_{\lambda, \lambda'} \int \frac{d\bm{k}}{(2\pi)^{2}}  \vert t_{\lambda, \lambda'}(\bm{k},\bm{k}')\vert^2 \nonumber \\
&& \times \int \frac{d \varepsilon}{2\pi}{\cal A}_{\lambda}(\bm{k},\varepsilon; \mu_{\rm B}) {\cal A}_{\lambda'}(\bm{k}',\varepsilon-\Delta \varepsilon_{\rm D}; \mu_{\rm T}) \nonumber \\
&& \times [n_{\rm F}(\varepsilon;\mu_{\rm B})-n_{\rm F}(\varepsilon -\Delta \varepsilon_{\rm D};\mu_{\rm T})]~,
\end{eqnarray}
where $N_{\rm f}=4$ is the number of fermion flavors in graphene and the wave vector $\bm{k}'$ in the top layer is fixed by momentum conservation to the value $\bm{k}' = \bm{k}-\Delta{\bm K}_{j}$.
(Any choice of $j=1,2,3$ is possible due to the three-fold rotational symmetry of the system.)
In Eq.~(\ref{eq2:current})
\begin{equation}
n_{\rm F}(\varepsilon; \mu) =\left\{ \exp{\left[\frac{\varepsilon- \mu}{k_{\rm B} T}\right]} + 1 \right\}^{-1}
\end{equation}
is the Fermi-Dirac distribution function at temperature $T$ and chemical potential $\mu$, while
\begin{equation}
{\cal A}_{\lambda}({\bm k},\varepsilon; \mu) = \frac{-2 \Sigma_\lambda''({\bm k},\varepsilon;\mu)}{[\varepsilon - \varepsilon_{{\bm k}, \lambda} - \Sigma_\lambda ' ({\bm k},\varepsilon;\mu)]^2+[\Sigma_\lambda '' ({\bm k},\varepsilon;\mu)]^2}
\end{equation}
is the spectral function of an interacting system of 2D massless Dirac fermions~\cite{polini_prb_2008}, here expressed in terms of the real $\Sigma_{\lambda} ' ({\bm k}, \varepsilon;\mu)$ and imaginary  
$\Sigma_{\lambda} '' ({\bm k}, \varepsilon;\mu)$ parts of the retarded quasiparticle self-energy $\Sigma_{\lambda}({\bm k}, \varepsilon;\mu)$.
The chemical potentials in the bottom and top layers are denoted by $\mu_{\rm B}$ and $\mu_{\rm T}$, respectively, and are measured with respect to the energy of the Dirac point in the corresponding layer.
The Dirac points of the top and bottom layers are offset by an energy $\Delta \varepsilon_{\rm D}$.

\begin{figure}
\begin{overpic}[width=\linewidth]{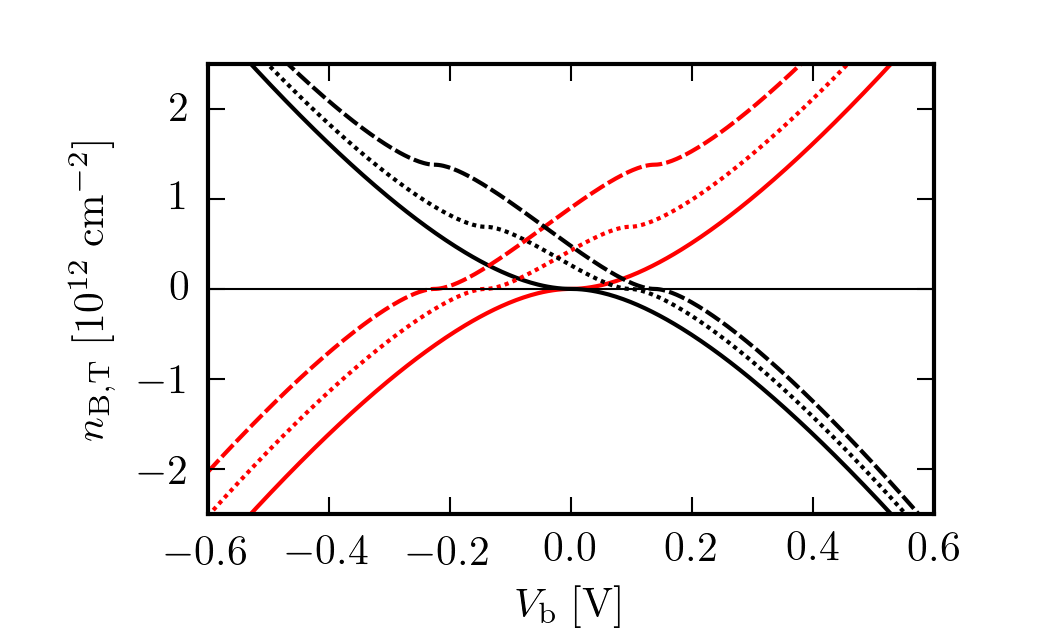}\put(1.8,55){(a)}\end{overpic}
\begin{overpic}[width=\linewidth]{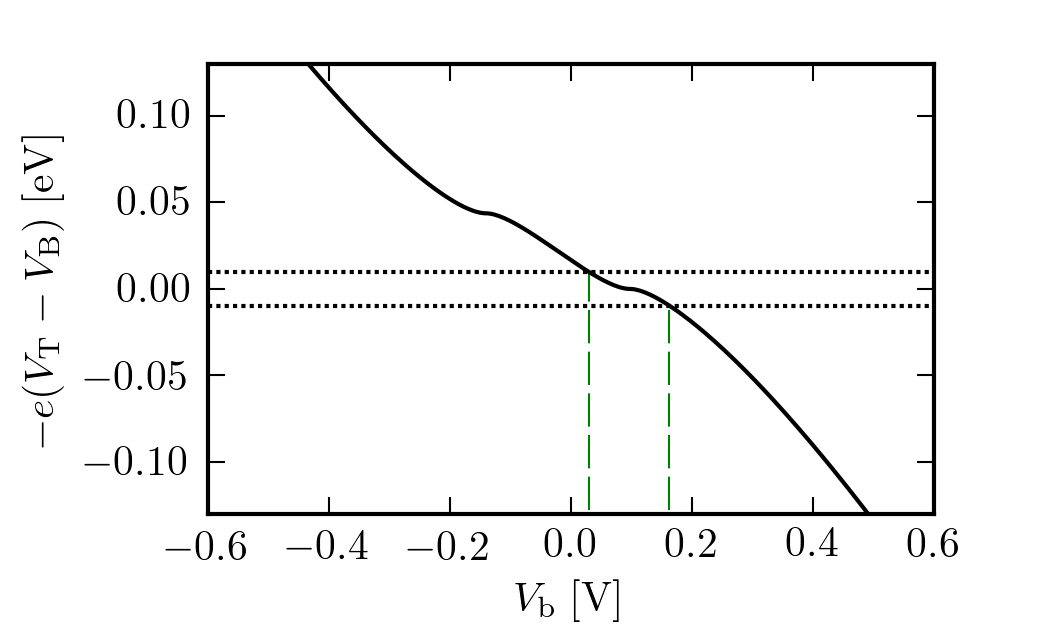}\put(1.8,55){(b)}\end{overpic}
\caption{\label{fig:electro}
(Color online)  Panel (a) The carrier densities $n_{\rm T}$ on the top (black lines) and $n_{\rm B}$ on the bottom (red lines) graphene layers are plotted as a function of the inter-layer bias $V_{\rm b}$, for different values of gate voltage: $V_{\rm G} = 0$ (solid lines), $V_{\rm G} = 10~{\rm V}$ (dotted lines), and $V_{\rm G} = 20~{\rm V}$ (dashed lines). Panel (b) The energy offset $\Delta \varepsilon_{\rm D} = -e(V_{\rm T}-V_{\rm B})$ (solid line) between the Dirac
points of the top and bottom layers is plotted as a function of inter-layer bias. 
Here, the misalignment angle is set at a small value, $\theta = 0.05^{\circ}$, and the gate voltage is set at $V_{\rm G} = 10~{\rm V}$. Long-dashed vertical lines mark the values of inter-layer bias voltage 
at which the collinearity condition~(\ref{eq:collinearity}) is met.}
\end{figure}

\section{Electrostatics and the quasiparticle spectral function}
\label{sect:details}

In this Section we summarize the two crucial ingredients that are required 
for the calculation of the tunneling current: i) electrostatic relations linking the chemical potentials $\mu_{\rm T}$ and $\mu_{\rm B}$ in the two layers with gate voltage $V_{\rm G}$ and inter-layer bias $V_{\rm b}$ and ii) details on the quasiparticle spectral function $A_{\lambda}(\bm{k},\varepsilon; \mu)$ and quasiparticle lifetime effects.

\subsection{Electrostatics}
\label{subsect:electrostatics}

For the sake of completeness, we here report a closed system of equations~\cite{mishchenko_naturenano_2014} relating the chemical potentials $\mu_{\rm B}$ and 
$\mu_{\rm T}$ and the energy offset $\Delta \varepsilon_{\rm D}$ to the gate voltage $V_{\rm G}$ and inter-layer bias $V_{\rm b}$. We remark that the chemical potential in each layer is measured with respect to the Dirac point of that layer.

The energy offset between top and bottom graphene layers is defined by 
\begin{equation}\label{eq:delta-epsilon-D}
\Delta \varepsilon_{\rm D} \equiv -e(V_{\rm T}- V_{\rm B})~, 
\end{equation}
where $V_{\rm B}$ ($V_{\rm T}$) is the magnitude of the electric potential at the bottom (top) layer.
Here, we assume that all quantities do not change in the ${\hat {\bm x}}$-${\hat {\bm y}}$ plane, i.e.~in the direction perpendicular to the ``growth'' direction of the van der Waals stack. 

The electro-chemical potential $\tilde{\mu}_{\rm B,T}$ in each graphene layer is given by the sum of the chemical potential and the electric potential energy, i.e.~$\tilde{\mu}_{\rm B,T} \equiv \mu_{\rm B,T}-e V_{\rm B,T}$.
The difference between the electro-chemical potentials of the top and bottom layers is due to the applied bias voltage, i.e.~$-eV_{\rm b}= \tilde{\mu}_{\rm T}- \tilde{\mu}_{\rm B}$.
Combining the above equations, we find the following electrostatic relation
\begin{equation}\label{eq:sys1}
-eV_{\rm b} = \mu_{\rm T} + \Delta \varepsilon_{\rm D}-\mu_{\rm B}~.
\end{equation}

A second electrostatic relation follows from the {\it charge neutrality condition}:
\begin{equation}\label{eq:neutrcharge}
n_{\rm B} +n_{\rm T} + n_{\rm G} = 0~,
\end{equation}
where $n_{\rm B}$, $n_{\rm T}$, and $n_{\rm G}$ are the charge densities on the bottom graphene layer, top graphene layer, and back gate, respectively.
We assume that both graphene layers have negligible residual doping.

We now relate these carrier densities to $V_{\rm G}$ and $V_{\rm b}$.
Using Gauss theorem, we find
\begin{eqnarray} \label{eq:gausslaw}
E_{1} & = & -e n_{\rm G}/(\epsilon_0 \epsilon_{\rm r}) \nonumber \\
E_{2}-E_{1} & = & -e n_{\rm B}/(\epsilon_0 \epsilon_{\rm r})~,
\end{eqnarray}
where 
$E_{1}$ is the magnitude of the electric field in the $\hat{\bm z}$ direction between the gate and bottom graphene layer, while $E_{2}$ is the magnitude of the electric field in the $\hat{\bm z}$ direction between the bottom and top graphene layers.
In Eq.~(\ref{eq:gausslaw}), $\epsilon_0$ is the vacuum permittivity, while $\epsilon_{\rm r}$ is an effective relative dielectric constant describing screening due to the dielectric materials surrounding the graphene layers. For sake of simplicity, we follow Ref.~\onlinecite{mishchenko_naturenano_2014} and take 
$\epsilon_{\rm r} = 4$. One can easily improve on this approximation by a more detailed electrostatic calculation that takes into account the uniaxial nature of hBN and thin-film effects (see e.g.~Ref.~\onlinecite{tomadin_prl_2015}).

The electric fields are related to the electric potentials on the graphene layers and on the gate by the relations
\begin{eqnarray} \label{eq:electricpotdef}
E_{1} & = & -(V_{\rm B} -V_{\rm G})/D \nonumber \\
E_{2} & = &  -(V_{\rm T} -V_{\rm B})/d~.
\end{eqnarray}

Finally, we can relate the chemical potential $\mu_{{\rm T}({\rm B})}$ to the carrier density $n_{{\rm T}({\rm B})}$ by using 
\begin{equation}\label{eq:mudef}
\mu \equiv \frac{\partial [n\varepsilon(n)]}{\partial n}~. 
\end{equation}
In Eq.~(\ref{eq:mudef}), $\varepsilon(n)$ is the ground-state energy per particle of the system of interacting fermions~\cite{barlas_prl_2007,asgari_annals_2014}, calculated independently in each layer. For example, to obtain $\mu_{\rm T}$ one needs to use Eq.~(\ref{eq:mudef}) with $n\to n_{\rm T}$ and $\varepsilon(n) \to \varepsilon_{\rm T} = \varepsilon_{\rm T}(n_{\rm T})$.

At temperatures $k_{\rm B} T\ll \varepsilon_{{\rm F}, {\rm B} ({\rm T})}$ and neglecting many-body exchange and correlation effects~\cite{barlas_prl_2007,asgari_annals_2014}, we can use the approximate relation
\begin{equation}\label{eq:chemdens}
\mu_{{\rm T} ({\rm B})} =\varepsilon_{{\rm F}, {\rm T}({\rm B})} \left [ 1 - \frac{\pi^{2}}{6}\left(\frac{k_{\rm B} T}{\varepsilon_{{\rm F}, {\rm T} ({\rm B})}}\right)^2 \right]~,
\end{equation} 
where $\varepsilon_{{\rm F}, {\rm T}({\rm B})} = \hbar v_{\rm F} \sqrt{4 \pi n_{{\rm T}({\rm B})}/N_{\rm f}}$ is the Fermi energy in each layer and $v_{\rm F} \sim 10^{6}~{\rm m}/{\rm s}$ is the graphene Fermi velocity.

Without loss of generality, we assume that the bottom layer is grounded, which implies $V_{\rm B} = 0$.
Eqs.~(\ref{eq:sys1}), (\ref{eq:neutrcharge}), (\ref{eq:gausslaw}), (\ref{eq:electricpotdef}), and~(\ref{eq:chemdens}) can 
be solved for the eight unknowns 
$n_{\rm B}$ $n_{\rm T}$, $n_{\rm G}$, $\mu_{\rm T}$, $\mu_{\rm B}$, $E_{1}$, $E_{2}$, and $V_{\rm T}$, as functions of the experimentally relevant parameters $V_{\rm b}$ and $V_{\rm G}$. Typical results are shown in Fig.~\ref{fig:electro}.

\subsection{The quasiparticle spectral function}
\label{subsect:quasiparticle}

In this Article we are interested in the impact of quasiparticle lifetime effects on the tunneling spectra of nearly-aligned graphene sheets. For the sake of simplicity, we use a Lorentzian approximation for the quasiparticle spectral function:
\begin{equation}\label{eq:approximate_spectral_function}
\mathcal{A_{\lambda}}(\bm{k},\varepsilon; \mu)= \frac{\hbar/ [\tau(\varepsilon_{\bm{k},\lambda}-\mu)]}{(\varepsilon-\varepsilon_{\bm{k},\lambda})^2+\lbrace \hbar/ [2 \tau(\varepsilon_{\bm{k},\lambda}-\mu) ] \rbrace^2 }~.
\end{equation}
In Eq.~(\ref{eq:approximate_spectral_function}), $\varepsilon_{\bm{k},\lambda}=\lambda \hbar v_{\rm F} |\bm{k}|$ is the Dirac band energy~\cite{kotov_rmp_2012} and
\begin{equation}\label{eq:Matthiessen}
\frac{\hbar}{\tau(\xi)} = \frac{\hbar}{\tau_{\rm ee}(\xi)} + \frac{\hbar}{\tau_{\rm s}}~.
\end{equation}
The quantity $\tau_{\rm ee}(\xi)$ is the lifetime of a quasiparticle of energy $\xi$ (measured from the chemical potential) and is related to the imaginary part of the retarded self-energy by the relation $[ \tau(\varepsilon_{{\bm k},\lambda} - \mu)]^{-1} = - 2\Sigma_{\lambda} '' ({\bm k}, \varepsilon_{{\bm k},\lambda};\mu)/\hbar$. In the spirit of Matthiessen's rule~\cite{hwang_prb_2008}, in Eq.~(\ref{eq:Matthiessen}) we have included a temperature-independent spectral width $\hbar /\tau_{\rm s}$ to take into account the effect of elastic scattering off of the static disorder potential on the quasiparticle lifetime. 

In the high-temperature $|\xi| \ll k_{\rm B} T$ limit, the expression for the decay rate $\hbar/ \tau_{\rm ee}(\xi)$ due to e-e interactions near the Fermi surface  is independent of $\xi$ and reads as following~\cite{polini_arxiv_2014,li_prb_2013}:
\begin{equation}\label{eq:taufinitet}
\frac{\hbar}{\tau_{\rm ee}(\xi)} = \frac{\pi}{4}\frac{(k_{\rm B} T)^2}{|\varepsilon_{\rm F}|}\ln \left (\frac{\Lambda}{k_{\rm B} T} \right )~,
\end{equation}
$\Lambda$ being a suitable cutoff~\cite{polini_arxiv_2014}. On the contrary, in the low-temperature $k_{\rm B} T \ll |\xi|$ limit the lifetime depends on the quasiparticle energy and is given by~\cite{polini_arxiv_2014,li_prb_2013} 
\begin{equation}\label{eq:tauzerot}
\frac{\hbar}{\tau_{\rm ee}(\xi)} = \frac{1}{4 \pi}\frac{\xi^2}{|\varepsilon_{\rm F}|}\ln \left ( \frac{\Lambda}{|\xi|} \right )~.
\end{equation}

The simple Lorentzian approximation (\ref{eq:approximate_spectral_function}), which has already been used e.g. in Ref.~\onlinecite{mishchenko_naturenano_2014} in the non-interacting 
$\tau_{\rm ee} \to \infty$ limit, can be transcended by employing the GW-RPA approximation~\cite{polini_prb_2008}. A study of these refinements on the spectral function and a detailed investigation of the role of graphene plasmons in the tunneling spectra~\cite{polini_prb_2008,principi_ssc_2012} is well beyond the scope of the present Article and will be discussed elsewhere.

\section{Numerical results and discussion}
\label{sect:numerics}

\begin{figure}
\begin{overpic}[width=\linewidth]{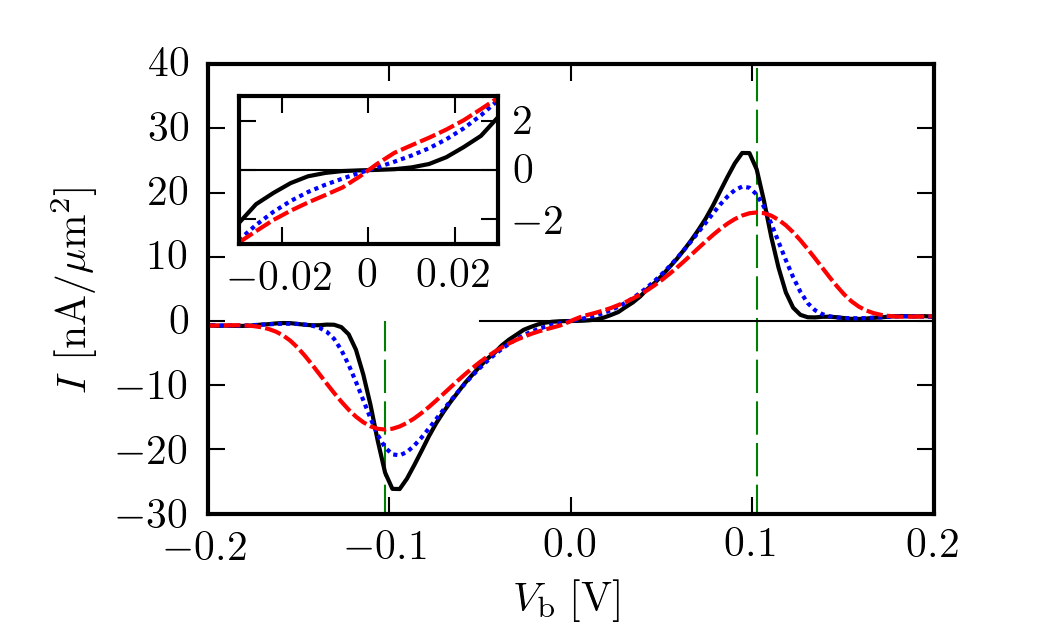}\put(2,55){(a)}\end{overpic}
\begin{overpic}[width=\linewidth]{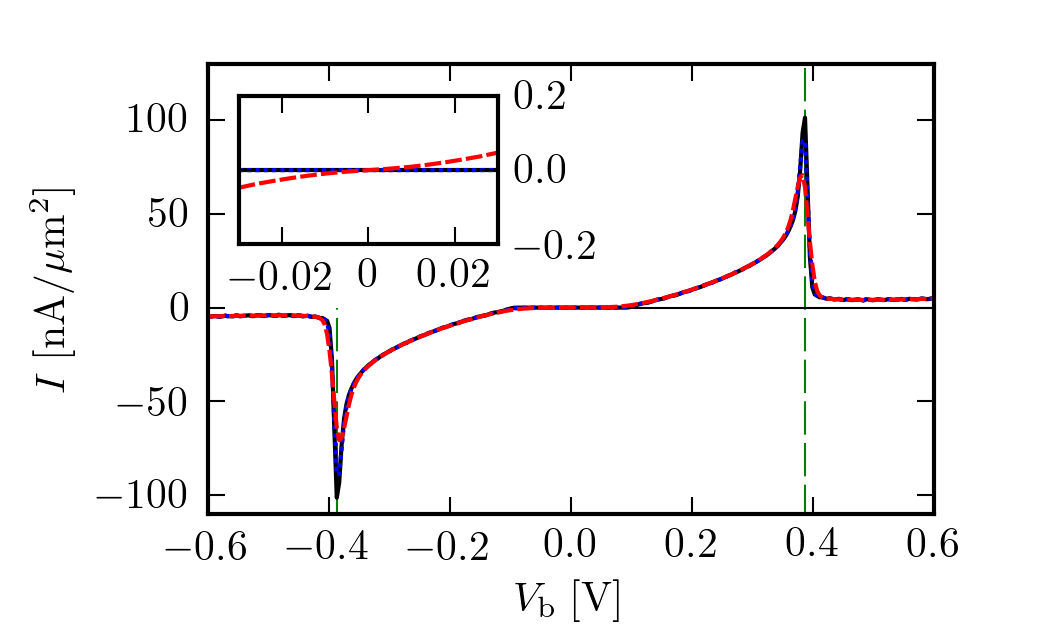}\put(2,55){(b)}\end{overpic}
\caption{\label{fig:deptemp}
(Color online) Current density $I$ as a function of the inter-layer bias voltage $V_{\rm b}$ for different values of temperature: $T = 10~{\rm K}$ (black solid line), $T = 45~{\rm K}$ (blue dotted line), and $T = 100~{\rm K}$ (red dashed line). Panel (a) Results for $\theta = 0.05^{\circ}$. Panel (b) Results for $\theta = 0.5^{\circ}$. 
In both panels, the inset shows a magnification of the curves around $V_{\rm b} = 0$. 
As in Fig.~\ref{fig:electro}(b), long-dashed vertical lines denote the value of $V_{\rm b}$ where the collinearity condition (\ref{eq:collinearity}) is met. All results in this figure have been obtained by setting $V_{\rm G} = 0$.}
\end{figure}

\begin{figure}
\includegraphics[width=\linewidth]{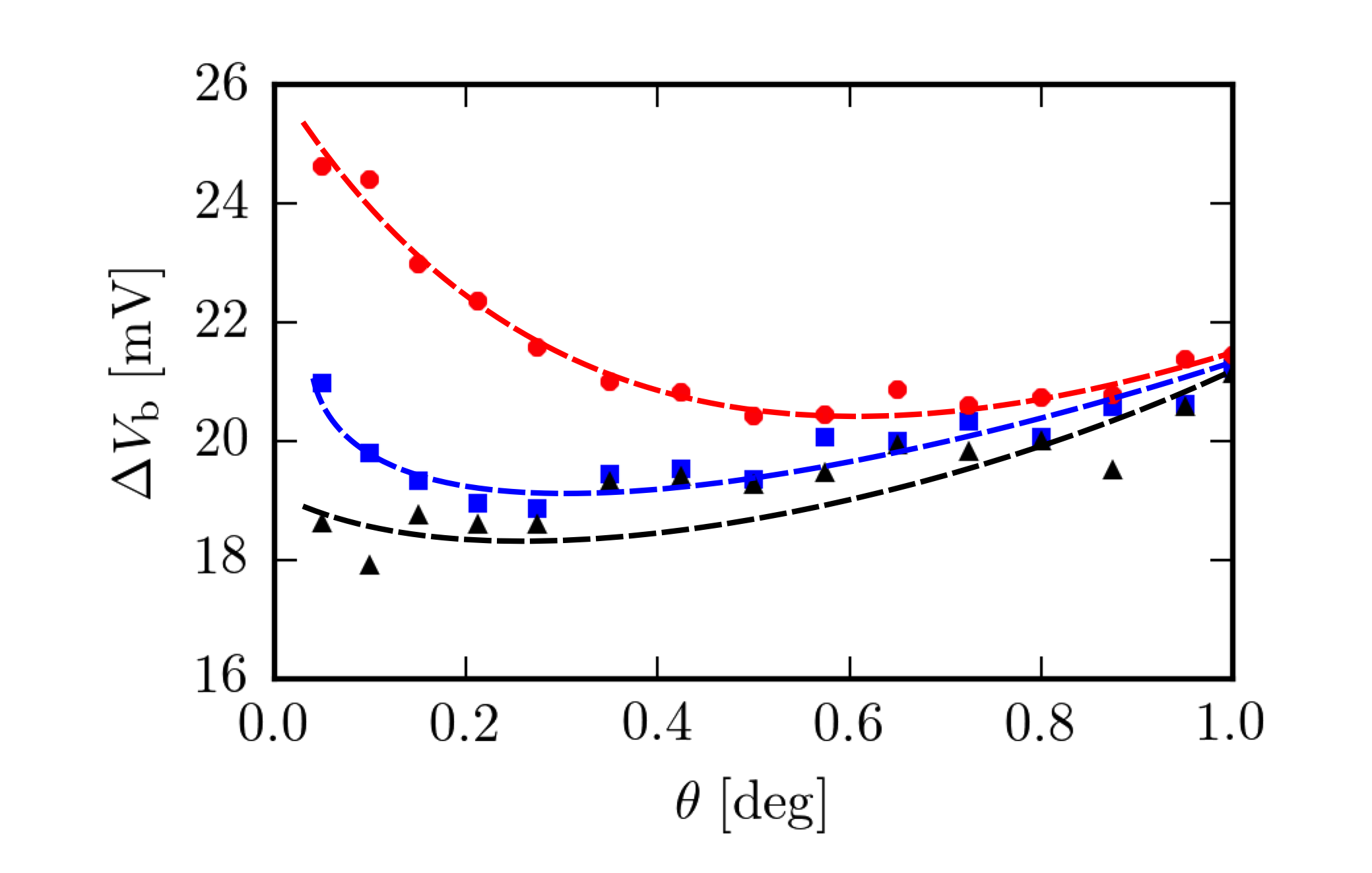}
\caption{\label{fig:broadening}
(Color online) Broadening $\Delta V_{\rm b}$ of the current density peak as a function of the misalignment angle $\theta$. Different sets of data refer to different values of temperature: $T = 10~{\rm K}$ (black triangles), $T = 45~{\rm K}$ (blue squares), and $T = 100~{\rm K}$ (red circles).
The dashed lines are guides to the eye. Data in this plot have been obtained by setting $V_{\rm G} = 0$. Quasiparticle lifetime effects emerge for small values of the misalignment angle.}
\end{figure}

We calculate the tunneling current by numerically performing the integrals in Eq.~(\ref{eq2:current}).
For the integration over the wave vector ${\bm k}$, we use a square mesh centered around the Dirac point, with maximum wave vector $k_{\rm max} = 2~{\rm nm}^{-1}$ and step $\Delta k = 4\times 10^{-2}~{\rm nm}^{-1}$. We have verified that the results do not change appreciably by using $k_{\rm max}$ up to $6~{\rm nm}^{-1}$. The energy mesh is symmetric and extends up to $\varepsilon_{\rm max} = 2~{\rm eV}$ with step $\Delta \varepsilon \lesssim 2\times 10^{-4}~{\rm eV}$.
In all numerical calculations we set $\Lambda = 3~{\rm eV}$, $D = 320~{\rm nm}$, and $\hbar /\tau_{\rm s} = 2~{\rm meV}$.
Finally, we set the hBN barrier thickness at $d = 1.4$ nm (approximately corresponding to $4$ hBN layers) and the effective coupling strength in Eq.~(\ref{eq:tunnham}) at $\gamma_{\rm eff} = 3~\mu{\rm eV}$. The latter choice is made to match the order of magnitude of the tunneling current measured experimentally~\cite{mishchenko_naturenano_2014}.

\begin{figure}
\begin{overpic}[width=\linewidth]{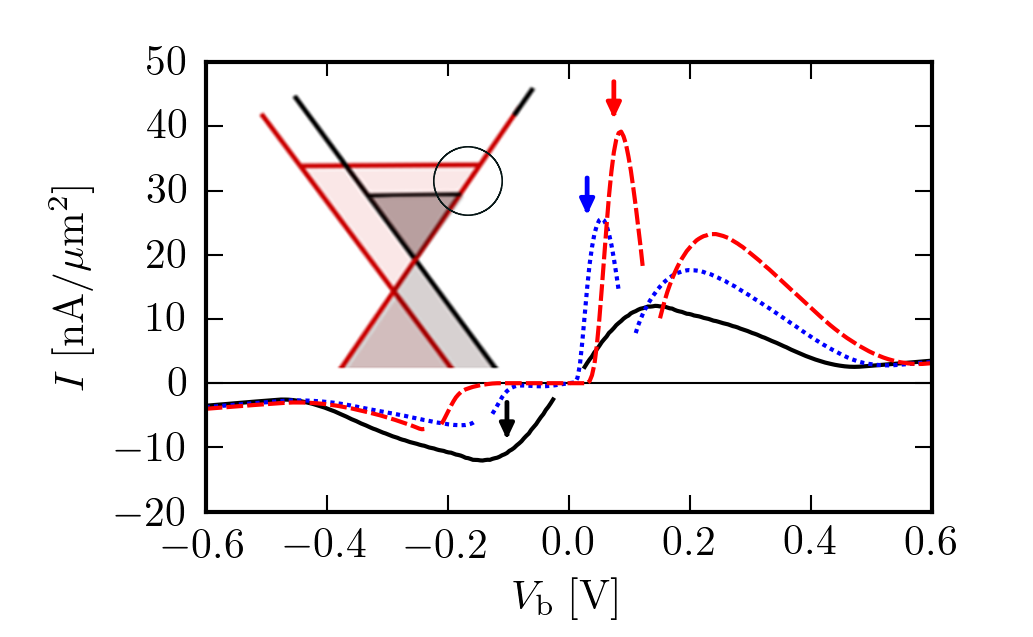}\put(2,55){(a)}\end{overpic}
\begin{overpic}[width=\linewidth]{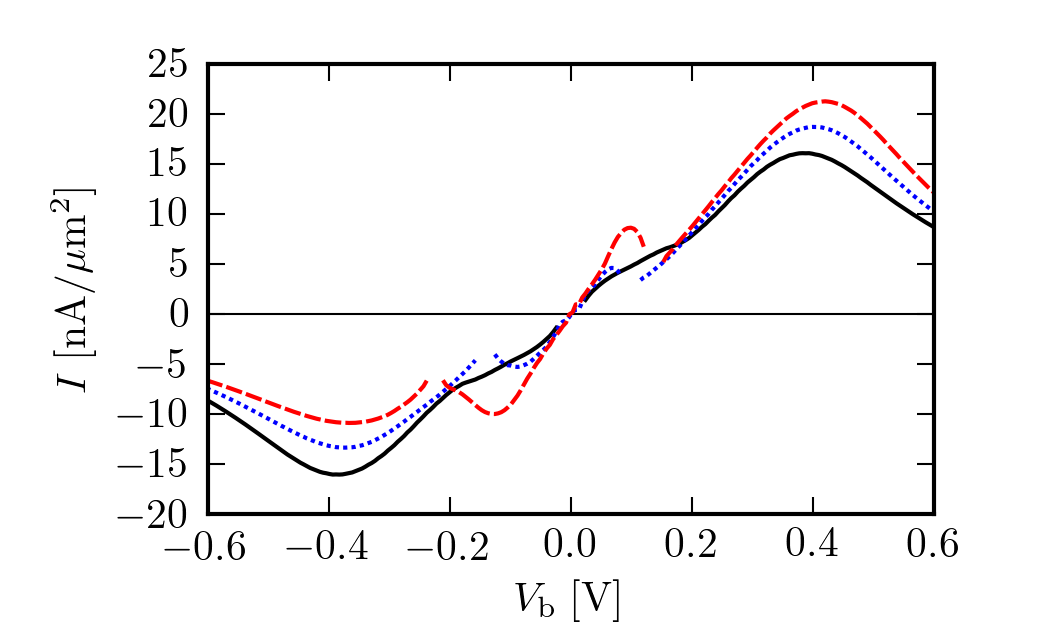}\put(2,55){(b)}\end{overpic}
\caption{\label{fig:gatedep}
(Color online) Current density $I$ as a function of the inter-layer bias voltage $V_{\rm b}$ for  
$T = 2~{\rm K}$ and different values of gate voltage: $V_{\rm G} = 0$ (black solid line), $V_{\rm G} = 10~{\rm V}$ (blue dotted line), and $V_{\rm G} = 20~{\rm V}$ (red dashed line). Panel (a) Results for $\theta = 0.05^{\circ}$. Panels (b) Results for $\theta = 0.5^{\circ}$. 
Tunneling curves are not shown for values of $V_{\rm b}$ such that 
$|n_{\rm T}|, |n_{\rm B}| \leq 10^{10}~{\rm cm}^{-2}$. 
In panel (a), arrows mark the values of $V_{\rm b}$ at which a peak due to the collinearity condition~(\ref{eq:collinearity}) with $\sigma = 1$ is expected.
Note that, for $\theta = 0.05^{\circ}$, peaks at  $V_{\rm G} = 10~{\rm V}$ and $20~{\rm V}$ appear at $V_{\rm b} > 0$. In this regime, states satisfying the resonant tunneling condition lie along the common directrix of the two Dirac cones and are close to {\it both} $\mu_{\rm T}$ and $\mu_{\rm B}$---see inset in panel (a).}
\end{figure}

Our main numerical results are summarized in Figs.~\ref{fig:deptemp}-\ref{fig:gatedep}. We clearly see that the current density as a function of bias voltage displays two peaks, which occur when the following condition is met~\cite{mishchenko_naturenano_2014}:
\begin{equation}\label{eq:collinearity}
\Delta \varepsilon_{\rm D} = \sigma \hbar v_{\rm F} |\Delta \bm{K}_{j}|,
\quad \sigma = \pm 1~.
\end{equation}
To visualize the geometric meaning of this condition, it is useful to represent the conical band structures of the two graphene layers on the same wave vector-energy plane $({\bm k},\varepsilon)$, with the Dirac points displaced horizontally by $\Delta {\bm K}_{j}$ and vertically by $\Delta \varepsilon_{\rm D}$.
Each point on the surface of a Dirac cone corresponds to a single-particle state on one of the two layers.
Because of energy and momentum conservation, electron tunneling is possible only between pairs of single-particle states, on opposite layers, which correspond to the {\it same} point on the plane $({\bm k},\varepsilon)$.
In other words, states which can undergo energy-conserving tunneling correspond to the intersection of  each layer's Dirac cone with the other layer's displaced Dirac cone.
The finite width of the spectral function relaxes energy conservation and broadens the region of $({\bm k},\varepsilon)$ space where the tunneling process has a non-vanishing probability to occur.

The condition (\ref{eq:collinearity}) with $\sigma = 1$ ($\sigma = -1$) corresponds to the situation in which the top layer's Dirac point falls on the bottom layer's upper (lower) Dirac cone.
These two cases correspond to tunneling between states close the Dirac point of the top layer and those in the conduction and valence band of the bottom layer, respectively.
In such configuration, the intersection between the two cones---which is in general an ellipse, a hyperbola, or a parabola---degenerates to a single line, such that all the wave vectors of states participating in the tunneling process are {\it collinear} to $\Delta {\bm K}_{j}$.
For this reason, we refer to (\ref{eq:collinearity}) as to the ``collinearity'' condition.
It is well known that, for 2D massless Dirac fermions, collinear scattering yields a divergent spectral density of electron-hole pairs (see, for example, Ref.~\onlinecite{polini_arxiv_2014} and references therein to earlier work) and ultrafast non-equilibrium dynamics of photo-excited carriers~\cite{tomadin_prb_2013,brida_naturecommun_2013}.

Peaks in the current density at collinearity are symmetric with respect to $V_{\rm b} = 0$ for $V_{\rm G} = 0$, as in Fig.~\ref{fig:deptemp}, while the current profile is asymmetric for finite values of $V_{\rm G}$, as in Fig.~\ref{fig:gatedep}.
The asymmetry between the two graphene layers is a consequence of the position of the gate layer.
The value of the inter-layer bias potential at which the collinearity condition is met is found as explained in Fig.~\ref{fig:electro}(b). Here, the dotted horizontal lines, displaying $\pm \hbar v_{\rm F} |\Delta \bm{K}_{j}|$, are intersected with the solid line, displaying $\Delta \varepsilon_{\rm D}$.
For large regions of parameter space, the peak corresponding to $\sigma = 1$ ($\sigma = -1$) appears at negative (positive) bias voltages.
However, at very small angles and sufficiently large $V_{\rm G}$, the collinearity condition with both $\sigma = \pm$ may be met at $V_{\rm b}>0$.

The tunneling current density at finite temperature and for vanishing gate voltage is  shown in Fig.~\ref{fig:deptemp}. Data in this figure have been obtained by using Eq.~(\ref{eq:taufinitet}) for the quasiparticle lifetime.
Peaks at collinearity are evident and located at bias voltages close to those predicted on the basis of the simple expression (\ref{eq:collinearity}).
Increasing temperature, the peaks become broader and drift to slightly larger absolute values of the bias potential. Moreover, the linear dependence of the current on the bias voltage around $V_{\rm b} = 0$ becomes steeper as temperature increases.
Comparing the current profiles for two different values of the misalignment angle $\theta$ in the two panels of Fig.~\ref{fig:deptemp}, we see that these effects are much more evident for small misalignment angles.
This behavior is due to the fact that, for large values of $\theta$, broadening of the current peak is dominated by lattice misalignment effects, while e-e interactions play the most important role 
in the condition of near-alignment.
Indeed, temperature affects the tunneling current  through the suppression of the quasiparticle lifetime $\tau_{\rm ee}$, i.e.~broadening of the spectral function.
A broader spectral function entails a more relaxed energy conservation in the tunneling processes, and thus the collinear peak widens around its zero-temperature, geometrically-deduced position.
Varying temperature has no effect on the current profile, if the quasiparticle lifetime is not affected by e-e interactions.
Our results thus show that the tunneling current at sufficiently small misalignment angles bear clear signatures of e-e interactions.
This is central result of this Article.

To quantify the role of e-e interactions, in Fig.~\ref{fig:broadening} we plot the broadening of the current peak as a function of the misalignment angle $\theta$ for various temperatures.
Since the current profile around the peak is not symmetric and extends to large values of the bias voltage $V_{\rm b}$, the definition of ``peak broadening'' is not obvious.
Therefore, we adopt an {\it ad hoc} definition to estimate how temperature affects the peak broadening.
We define the broadening as the standard deviation $\Delta V_{\rm b} = \langle [V_{\rm b} - \langle V_{\rm b} \rangle]^{2} \rangle^{1/2}$, where the average
\begin{equation}
\langle X \rangle = \frac{ \int_{V_{{\rm b},1}}^{V_{{\rm b},2}} d V_{\rm b} ~ X(V_{\rm b}) I(V_{\rm b}) }{ \int_{V_{{\rm b},1}}^{V_{{\rm b},2}} d V_{\rm b} ~ I(V_{\rm b}) }
\end{equation}
is defined with respect to the current profile.
The extremes of integration $V_{{\rm b},1}$, $V_{{\rm b},2}$ are symmetric around the peak position $V_{{\rm b},{\rm peak}}$ with a total extent $V_{{\rm b},2} - V_{{\rm b},1} = 100~{\rm meV}$.
Fig.~\ref{fig:broadening} shows that the broadening of the current peak depends on temperature---a clear signature of e-e interactions. However, the temperature dependence is weak at misalignment angles $\theta \gtrsim 0.5^{\circ}$ (where the tunneling current away from collinearity is suppressed by lattice misalignment) and stronger at $\theta \lesssim 0.5^{\circ}$ (where the effect of e-e interactions becomes more important).

At low temperatures, a further signature of the electron spectral properties is found by studying the profile of the tunneling current as a function of gate voltage. This is shown in Fig.~\ref{fig:gatedep}.
Data in this figure have been obtained by using Eq.~(\ref{eq:tauzerot}) for the quasiparticle lifetime.
We have decided not to calculate the value of the current for ranges of $V_{\rm b}$ such that 
bottom- or top-layer carrier densities are smaller than $10^{10}~{\rm cm}^{-2}$. This is because 
the normal Fermi liquid expression~(\ref{eq:tauzerot}) for the quasiparticle decay rate is not justified near the charge neutrality point.
In these regions, the derivative of the carrier density of either layer with respect to $V_{\rm b}$ vanishes (see Fig.~\ref{fig:electro}).
As a consequence, the differential conductance $dI/dV_{\rm b}$ is nearly zero, as observed experimentally~\cite{mishchenko_naturenano_2014,britnell_naturecommun_2013}.

We observe that for small misalignment angles and large gate voltages the peak corresponding to the collinearity condition with $\sigma = 1$ [indicated by arrows in Fig.~\ref{fig:gatedep}(a)] is located at $V_{\rm b} > 0$. In this regime, the height of the peak 
is very sensitive to disorder and increases as the residual spectral width $\hbar/\tau_{\rm s}$ decreases. This is because the most important contribution to the energy integral in Eq.~(\ref{eq2:current})---due to collinearity---arises from a region in the $({\bm k},\varepsilon)$ plane where $\xi$ is small [see inset in Fig.~\ref{fig:gatedep}(a)].
That is, the dominant contribution to the tunneling current comes from particles tunneling from the neighborhood of the chemical potential in one cone to the neighborhood of the chemical potential in the other cone.
In this case, the e-e contribution to the quasiparticle lifetime tends to zero, as in all Fermi liquids, so that both the initial and final states involved in the tunneling process are long-lived and the tunneling probability is enhanced.
The finite height of the current peak is determined by the the residual spectral width due to disorder.
Similarly to the effect of e-e interactions at finite temperature, the effect of the residual spectral width is suppressed at larger misalignment angles [see Fig.~\ref{fig:gatedep}(b)], where the width and height of the current peaks is rather insensitive to the value of gate voltage.

\section{Summary}
\label{sect:summary}

In this Article we have presented a theory of the tunneling characteristics between misaligned graphene layers, which takes into account the spectral properties of the tunneling electrons.
We have taken into account quasiparticle lifetime effects into the quasiparticle spectral function by treating on an equal footing electron-electron interactions and elastic scattering off of the static disorder potential.
Effects of electron-electron interactions on the quasiparticle lifetime are considered separately at finite (Figs.~\ref{fig:deptemp}-\ref{fig:broadening}) and very low (Fig.~\ref{fig:gatedep}) temperatures. In both cases, we study the interplay between the misalignment angle and the quasiparticle lifetime.

The profile of the tunneling current as a function of the bias voltage is characterized by peaks which originate from the enhanced tunneling probability between electronic states with collinear wave vectors in the two layers. Due to electron-electron interactions, the broadening of these peaks depends on temperature at small misalignment angles. In this regime, comparing experimental data with our theoretical results enables measurements 
of the quasiparticle lifetime $\tau_{\rm ee}$ in a vertical transport experiment.
At very low temperatures, instead, by tuning the gate voltage, it is possible to reach a regime in which the height of one current peak is entirely determined by the quasiparticle lifetime due to elastic scattering.
Both effects disappear when the misalignment angle $\theta$ is larger than $0.5^{\circ}$-$1^{\circ}$, because, in this case, the width of the current peaks is dominated by the non-conservation of in-plane wave vector during the tunneling process.

Measurements of $\tau_{\rm ee}$ can be compared with many-body theory calculations~\cite{polini_arxiv_2014,li_prb_2013,principi_arxiv_2015} and are important to assess the region of parameter space (carrier density and temperature) where transport in massless Dirac fermion fluids can be described by hydrodynamic theory~\cite{bandurin_arxiv_2015,torre_prb_2015}.

\acknowledgments
K.A.G.B.~acknowledges useful discussions with J.R.~Wallbank, P.~D'Amico, and G.~Borghi. 
This work was supported by the EC under the Graphene Flagship program (contract no.~CNECT-ICT-604391) and MIUR through the program ``Progetti Premiali 2012'' - Project ``ABNANOTECH''.
We have made use of free software~\cite{python}.

\end{document}